\documentstyle[multicol,epsf,aps]{revtex}
\begin{document}
\title{Spontaneous spirals in vibrated granular chains}
\author{R.~E.~Ecke$^{1,2}$, Z.~A.~Daya$^{1,2}$, M.~K.~Rivera$^{1,2}$,
  and E.~Ben-Naim$^{1,3}$}
\address{${^1}$Condensed Matter \&
Thermal Physics Group,${^2}$Center for Nonlinear Studies, ${^3}$Theoretical Division}
\address{Los Alamos National Laboratory, Los Alamos, NM 87545}
\maketitle
\begin{abstract}

We present experimental measurements on the spontaneous formation of
compact spiral structures in vertically-vibrated granular
chains. Under weak vibration, when the chain is quasi two-dimensional
and self-avoiding, spiral structures emerge from generic initial
configurations. We compare geometrical characteristics of the spiral
with that of an ideal tight spiral. Globally, the spiral undergoes a
slow rotation such that to keep itself wound, while internally, fast
vibrational modes are excited along the backbone with transverse
oscillations dominating over longitudinal ones. Spirals have an
extremely small volume in phase space, and hence, their formation
demonstrates how nonequilibrium dynamics can result in a nonuniform
sampling of  phase space.
  
\end{abstract}

\begin{multicols}{2}

\section{Introduction}
\label{intro}
Ratchet mechanisms are thought to play an important role in producing
net transport or motion by asymmetrically absorbing energy from a
stochastic source
\cite{astumian_97,jap_97,astumian_94,bartussek_94}. Such mechanisms
should apply to systems far from equilibrium as well, specifically
with respect to the formation of patterns or structures. In other
words, spontaneously generated patterns might arise from a stochastic
forcing.  Shaken granular media are natural systems for investigating
possible ratchet effects in nonequilibrium systems
\cite{vicsek_98}. Mechanical vibration is athermal and coupled with
the dissipative collisions results in effectively a stochastic forcing
often manifested by anomalous velocity
fluctuations\cite{kudrolli_97,rouyer_00,olafsen_99}.

Recently, we have explored a chain of coupled hollow spheres attached
by rods and vibrated vertically on a horizontal surface
\cite{ebn_01,hastings_02}. The vertical vibration pumps energy into
the chain and, because the collisions with the plate are typically
asynchronous with the drive, the chain develops irregular vertical
structure. Collisions between balls on the chain generate a noisy
dynamics of lateral motions so that viewed from above the chain
appears as effectively two-dimensional. Further, when the vibration is
weak so that the vertical displacement of the chain is less than one
ball diameter, the chain cannot cross itself and is self-avoiding.

Previous experiments suggest that vibration induces strong
fluctuations which in turn cause the chain to explore all regions of
the available phase space, similar to the ergodicity principle in
equilibrium statistical mechanics.  Statistical properties of the
unraveling times of knots is consistent with random thermal motion of
the crossing points constituting the knot \cite{ebn_01}. Moreover,
two-loop figure-eight configurations in a ring chain prefer
states with maximal disparity in the size of the loops,
consistent with uniform sampling of the microscopic configuration
space \cite{hastings_02}.  Further, the statistical properties of free
and closed chains \cite{DRBE_02,prentis_02} are reminiscent of the
properties of ideal polymers in equilibrium systems
\cite{deGennes_79,doi_86} for such quantities as, for example, the
radius of gyration.

\begin{figure}
\narrowtext
\epsfxsize=5.7cm
\centerline{\epsffile{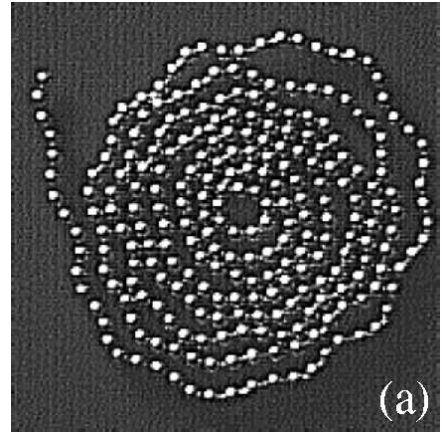}}
\epsfxsize=5.7cm
\centerline{\epsffile{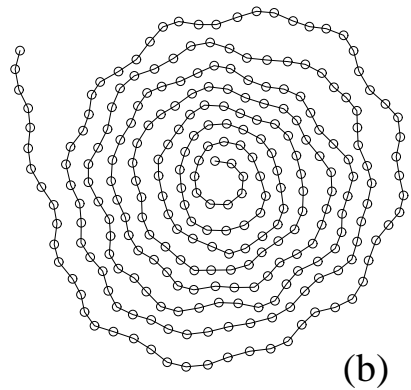}}
\caption{The raw image (a) and the reconstructed spiral (b).}
\label{image}
\end{figure}

One would expect on the basis of these previous results that a free
chain weakly excited so that it cannot cross itself would explore
phase space uniformly.  Such is not the case. Instead the chain may
form a compact spiral, see Figure~\ref{image}a, starting from some
generic initial condition. From the available phase space, the system
condenses into a very small region, namely the spiral state.  In this
paper, we present a characterization of the spiral state. We start
with a short description of the experimental apparatus, followed by
discussions of the formation, geometry and dynamics of the spirals.

\vspace{-3mm}
 
\section{Experimental Set-Up}
\label{expt}
The experiment consists of a vertically-vibrated anodized-aluminum
plate driven harmonically by an electromagnetic shaker. The plate was
circular with a diameter of $11.50$ inches. It was coupled to the
shaker by a shaft that occupied the bore of a precise square-section
linear air bearing. As a result, the plate's vertical vibration was
accurately sinusoidal with amplitude $A$ and angular velocity $\omega
= 2\pi f$ where $f$ is the driving frequency. In addition, the
square-section of the bearing ensured that the plate did not
experience rotational torques. An acrylic lateral boundary confined
the chain to the plate.  The instantaneous acceleration of the plate,
measured with an accelerometer, was maintained at constant value of
the dimensionless peak acceleration $\Gamma = a/g = \omega^2 A/g$ to
within 2\%.

The commercially-available ball chain consisted of $N$ hollow,
approximately spherical shells that were connected by dumb-bell shaped
rods. The chains were nickel-plated steel with a nominal bead diameter
$d={3\over32}$ inches with a variation of about $1\%$. The rods had a
diameter $0.22d$ and constrained the inter-bead separation $1 \leq b
\leq 1.4$ measured in units of the bead diameter. The maximal angle
between two consecutive rods was $\approx 50^\circ$ so that the smallest
circle that could be made from the chain had $n=8$ beads.  The ratio
of the diameter of the available plate to the bead diameter is about
$110$, and boundary effects were negligible.

In our experiments, a ball chain was placed onto the vibrating plate.
The relevant driving conditions are the frequency, $f$, and the
acceleration, $\Gamma$. We restricted our attention to weak vibration,
$\Gamma<2$, where the geometry is quasi two-dimensional.  The chain is
initiated in a random configuration subject only to the condition that
it did not intersect itself. In these weak driving conditions, the
chain remains close to the plate surface, free of self-intersections.
 
We obtained digital images of the chain using a high-resolution (1024
x 1024 pixels) CCD camera. Data sets were acquired at frame rates of
$1$~Hz to observe the slow global dynamics and at $250$~Hz to
characterize internal vibrational modes.  Using standard image
analysis procedures, we extracted positions of the bead centers. These
were then ordered to recover the locus of the spiral. For more details
see \cite{hastings_02}. A typical image and chain reconstruction are
shown in Figure~\ref{image}.

\section{Spiral formation}

We find experimentally that sufficiently long chains spontaneously
organize into spirals in a narrow range of accelerations, $1.70 \leq
\Gamma \leq 1.85$, and frequencies $10 \leq f \leq 25$~Hz. Starting
from a linear chain, a tight spiral core nucleates (either clockwise
or counter clockwise) at one end of the chain. This core rotates, and
as a result the chain winds into a spiral. The formation of compact
spirals from a nucleated core typically takes $3-7$ minutes for a
chain with $N=256$ beads. However, the nucleation of the spiral core
from a linear chain is a random event occurring on the scale of $30$
minutes. Once formed, the spiral was stable.

The formation of spirals is more sensitive to variations in the
acceleration rather than in the frequency, similar to pattern
formation and clustering in vibrated granular layers
\cite{umbanhower_96,olafsen_98}. When the acceleration is too large,
the chain exhibits strong excitations that quickly destroy the
spiral. On the other hand, when the acceleration is too small, lateral
chain motion is insufficient to nucleate the spiral core.

Interestingly, for a given chain, the spiral always nucleates with its
core at the same end. Moreover, artificially formed spirals with their
core at the opposite end are unstable. The mechanism responsible for
this symmetry breaking are minute imperfections in size, shape, and
weight of the extremal beads. Asymmetry underlies numerous transport
phenomena in physical and biological systems, for example, Brownian
motors\cite{astumian_94}.  Typically, asymmetry in the environment
coupled with oscillatory driving leads to ratchet motion. While
similar in spirit, granular chains are different in that the asymmetry
is in the transported object itself. Such a transport mechanism may be
relevant in biological systems, for example, it can be realized via
defects in the macromolecules. 

\section{Spiral Geometry}
\label{results}

The asymmetries are important in the nucleation of the spiral core and
its evolution to a compact rotating state. Since we are unable to
systematically control these asymmetries, we do not consider here the
transient formation stage, but rather, we focus on the characteristics
of the formed spiral. As a reference, we use a ideal tight
spiral. Consider such a spiral formed by a flexible rope. Without loss
of generality, its diameter is set to unity. A point along the spiral
is (over)specified by its arc length $s$, radius $r$, and winding
angle $\theta$.  It is convenient to think of the spiral as composed
of concentric tori.  In such an arrangement, the addition of the
outermost torus increases the angle by $\Delta\theta=2\pi$ and the arc
length by $\Delta s=2 \pi r$. Since $\Delta r = 1$ we have $ds/dr = 2
\pi r$ and $ds/d\theta = r$. It follows that $r(s)=\sqrt{s/\pi}$ and
$\theta(s)=\sqrt{4\pi s}$.

Consider now a bead chain.  The bead number $n$ is a natural
measure of the arc length, and assuming an average separation of
$\langle b\rangle$, one has $s(n)=n\langle b\rangle$. Moreover, the
distance between the $n$th bead and the $1$st bead, $r(n)$, is used as
a measure of the radius, since the spiral center is not uniquely
defined. The $(s,\theta)$ coordinates of the first bead are
$(1,0)$. Substituting $s(n)=n\langle b\rangle$ in the above
expressions for $\theta(s)$ and $r(s)$, and using the law of cosines
yields an expression for the winding angle and distance to the core as
a function of the bead number:
\begin{eqnarray}
\label{thetaeqn}
\theta(n)&=&\sqrt{4 \pi  n\langle b \rangle},\\
\label{reqn}
r(n)&=&\sqrt{1+{n\langle b \rangle\over \pi}-
2\sqrt{{n\langle b \rangle\over \pi}}\cos\theta(n)}.
\end{eqnarray}
Of course, these expressions apply to sufficiently large spirals and
$n\gg 1$.

We have measured the average angle and distance as a function of the
bead number from over $500$ frames. In these experiments and for the
remainder of this paper we have $N=256$ beads, $f=16$~Hz and
$\Gamma=1.77$. In reconstructing the spiral from the image our
ordering was such that the innermost core bead corresponds to $n=1$.

\begin{figure}
\narrowtext
\epsfxsize=6cm
\centerline{\epsffile{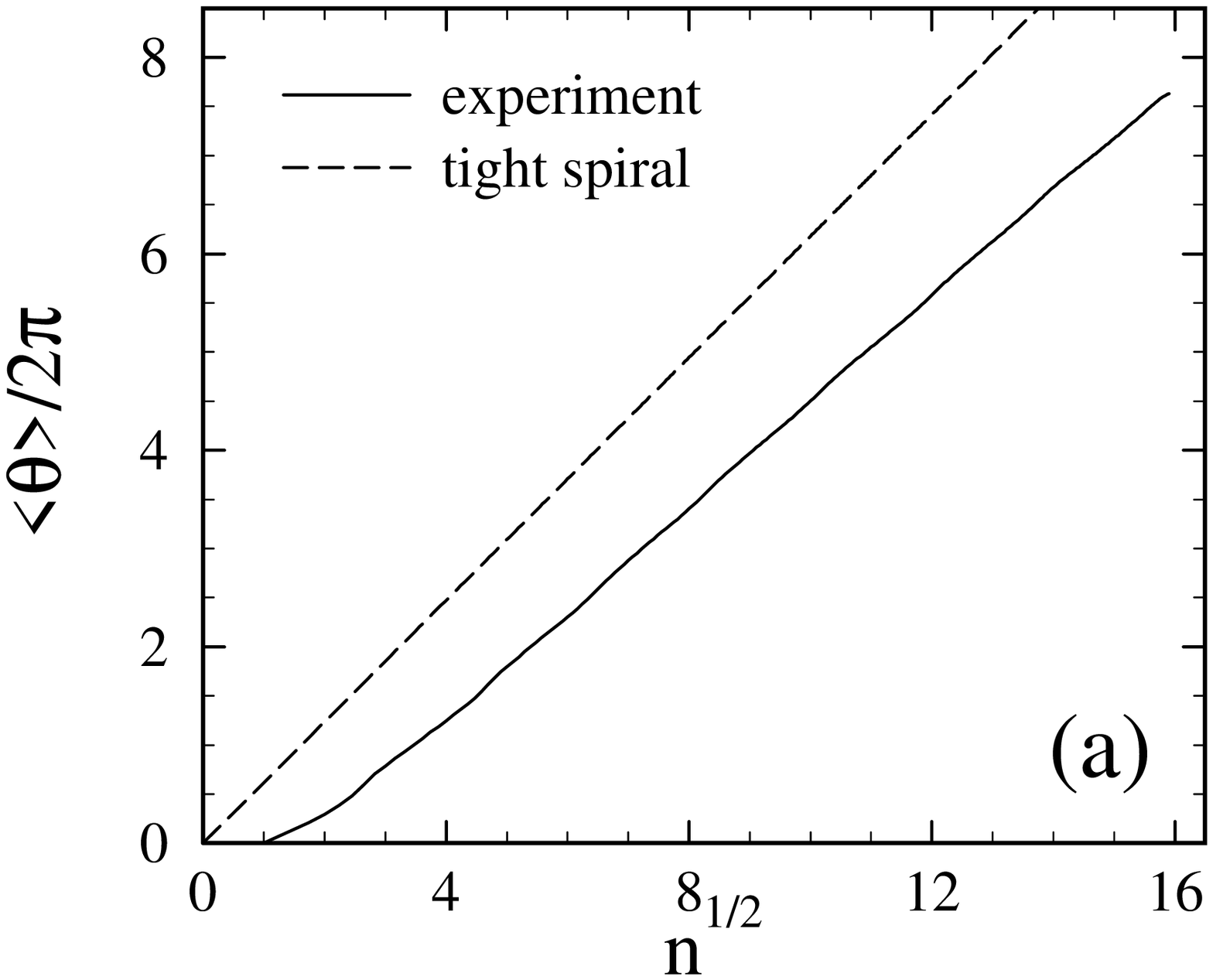}}
\epsfxsize=6cm
\centerline{\epsffile{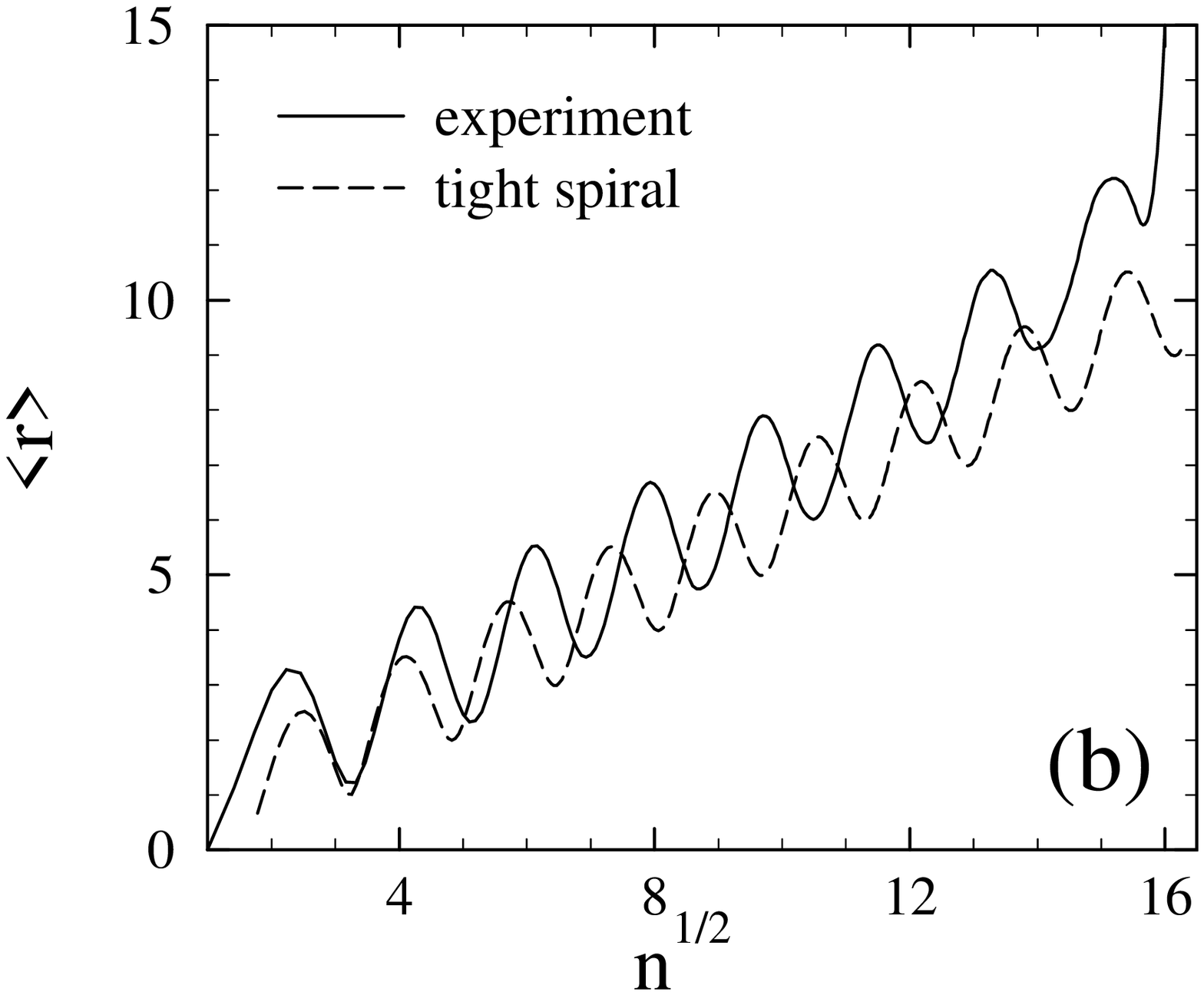}}
\epsfxsize=6cm
\centerline{\epsffile{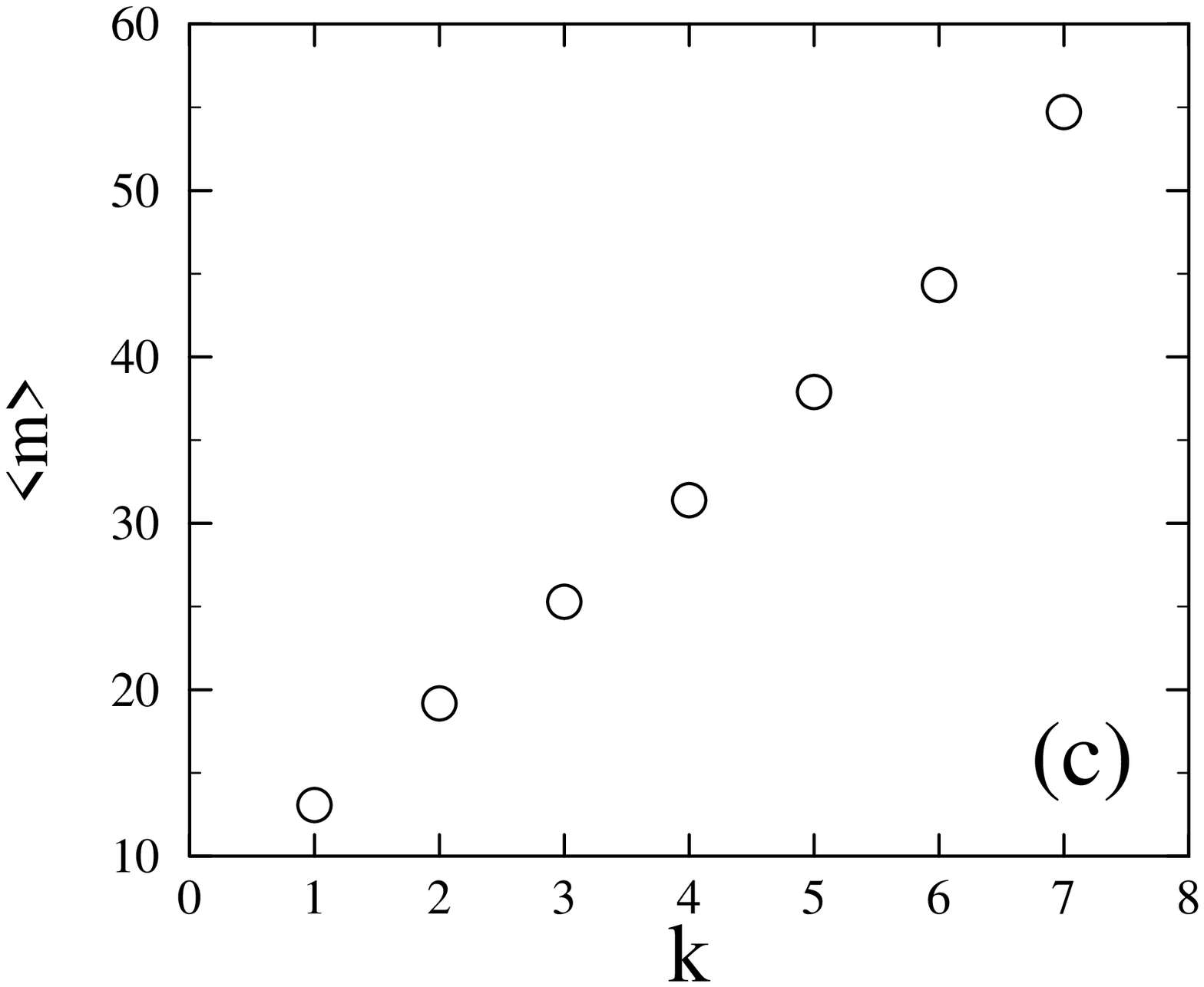}}
\caption{The mean angle $\langle \theta \rangle/2\pi$ (a) and distance
  $\langle r \rangle$ (b) as a
  function of $\sqrt{n}$. In (c) the mean number of beads $\langle m
  \rangle$ per turn is plotted vs the turn number $k$.}
\label{geometry}
\end{figure}
\vspace{-3mm}

In Figures~\ref{geometry}a and b we have plotted the averages $\langle
\theta \rangle (n)/2\pi$ and $\langle r \rangle (n)$ versus $\sqrt n$
respectively. For comparison, we also show the tight spiral angle and
distance calculated from Equations~(\ref{thetaeqn}) and
(\ref{reqn}). Here, the average separation $\langle b\rangle = 1.2$
was measured directly, as discussed below. The slower (faster) linear
growth of $\langle \theta \rangle (n)$ ($\langle r \rangle (n)$)
corresponds to the spiral becoming less tight the further one goes
from the compact core. Nevertheless, the gross features such as the
overall linear growth of $\langle r(n)\rangle$ with $\sqrt{n}$, and
the periodic corrections, with a period that is fixed in $\sqrt{n}$,
indicate that the spiral is approximately tight.

The periodic oscillations of $\langle r(n)\rangle$ allows for a
characterization of the number of beads in a given turn of the
spiral. A single turn is contained between two consecutive minima in
Figure~\ref{geometry}b. As shown in Figure~\ref{geometry}c, the mean
number of beads per turn grows linearly with the loop index,
consistent with the expected behavior for concentric tori. The
deviation from linearity for the last turn indicates that the spiral
has a straight tail. From this measurement, the average size of the
tail is estimated at $10$, consistent with direct observation. We also
find that different sets of $500$ images, corresponding to
independently formed spirals, are characterized by quantitatively
similar mean angle, distance and number of beads per turn as in
Figure~\ref{geometry}.

When comparing with a tight spiral, we implicitly assumed that the
separation between two beads is constant. Direct measurement shows that
this is a reasonable assumption as variations in $\langle b\rangle$ are
of the order of 10\%. However, this deviation has a clear trend. The
separation increases as a function of the bead number and it saturates
at a value of $\langle b\rangle\approx 1.24$ for $n>150$.

\begin{figure}
\narrowtext
\epsfxsize=6.4cm
\centerline{\epsffile{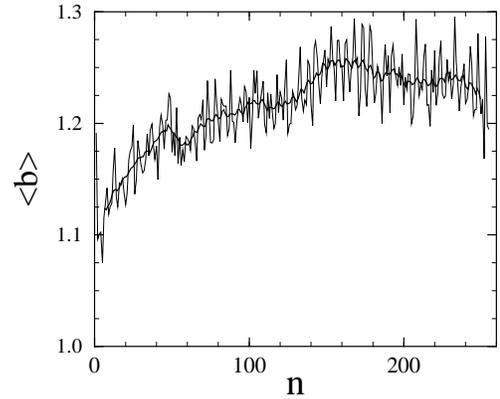}}
\caption{The average bead separation $\langle b\rangle$ (measured in
units of bead diameter $d$) as a function of the bead number $n$. The
thick line is a $12$-bead running average.}
\label{distance}
\end{figure}

\section{Spiral dynamics}

\begin{figure}
\narrowtext
\epsfxsize=6.1cm
\centerline{\epsffile{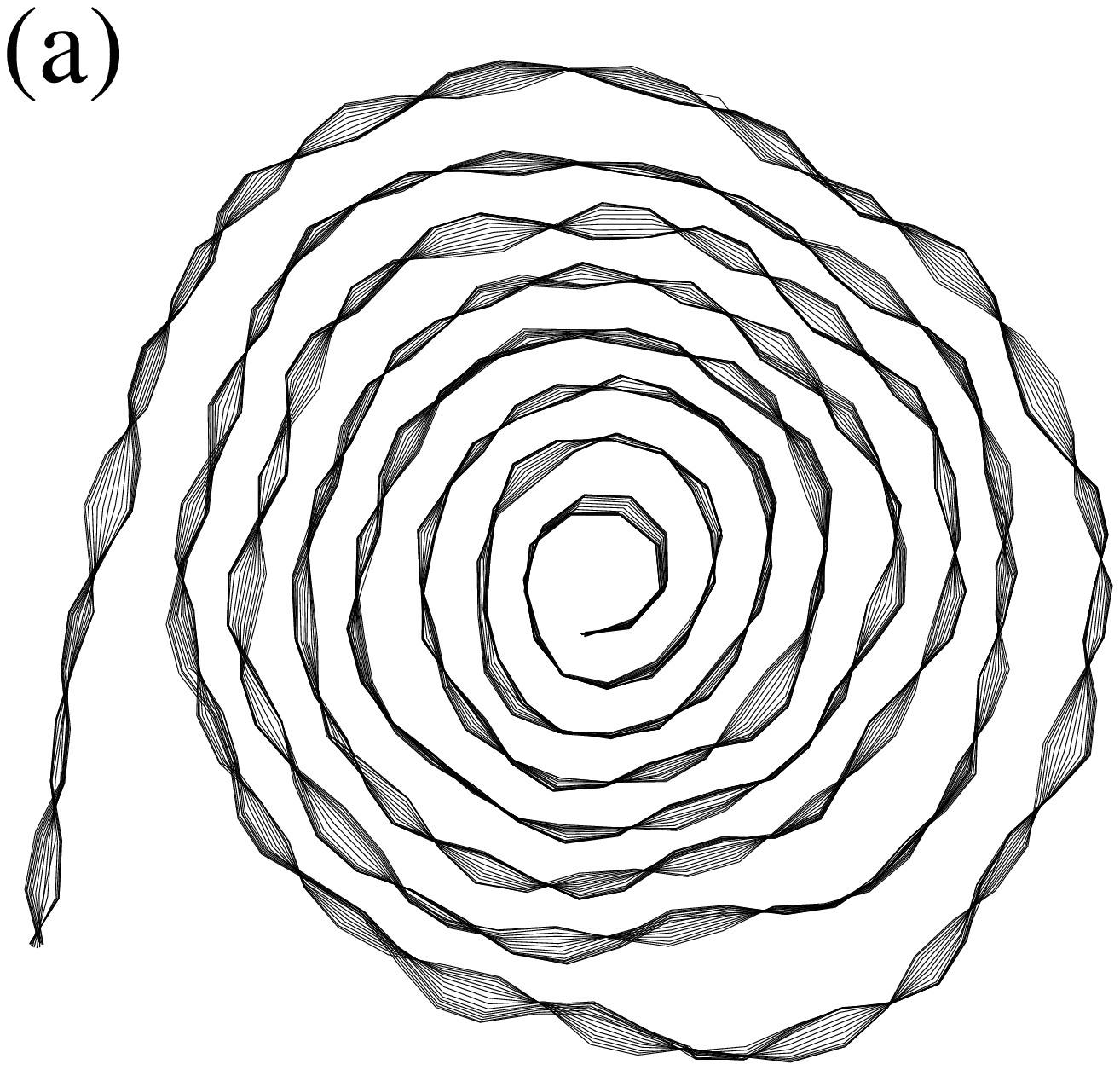}}
\epsfxsize=6.1cm
\centerline{\epsffile{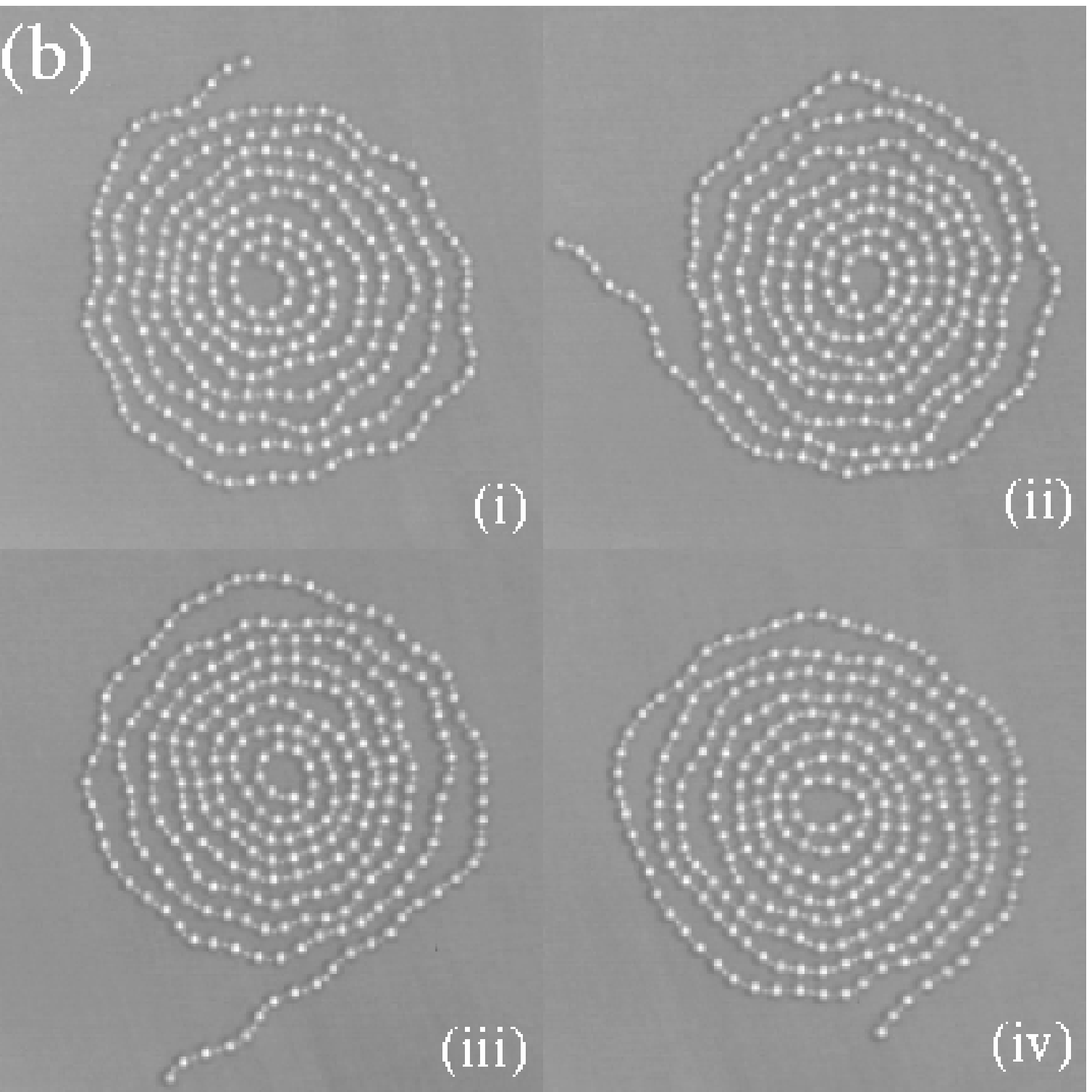}}
\caption{Sixteen consecutive reconstructions (a) of a spiral during 1
  oscillation showing the fast transverse wave modes. Images with a
  separation of 15 seconds (b) illustrating the slow rotation that keeps the spiral wound.}
\label{dynamics}
\end{figure}

Once formed, the vibrating spiral has dynamics on fast and slow time
scales. In Figure~\ref{dynamics}a, we have plotted an overlay of $16$
consecutive reconstructions of the spiral obtained at a sampling rate
of $250$~Hz. The $16$ images correspond to approximately $1$
oscillation of the plate. The lateral wave modes visible here
constitute in part the fast dynamics of the spiral. What is not
obvious in Figure~\ref{dynamics}a is the large number of collisions and
the constrained motion that effectively gives rise to the transverse
modes. Approximately $4$ beads separate the nodes of these modes. On a
much longer time scale we find that the spiral undergoes axial rotation.

\vspace{-3mm}

Four snapshots of a spiral with a separation of $15$ seconds or $240$
oscillations between snapshots is shown in Figure~\ref{dynamics}b. The
spiral rotates so as to keep tightly wound at a constant rate of about
one revolution per minute. On shorter time scales the rotation is not
constant, and is better described as a ratcheting motion with the
spiral core moving back and forth. The asymmetry between the winding
and unwinding motion of the core is transmitted along the spiral
length and over long time scales produces the global rotation.

\begin{figure}
\narrowtext
\epsfxsize=6.4cm
\centerline{\epsffile{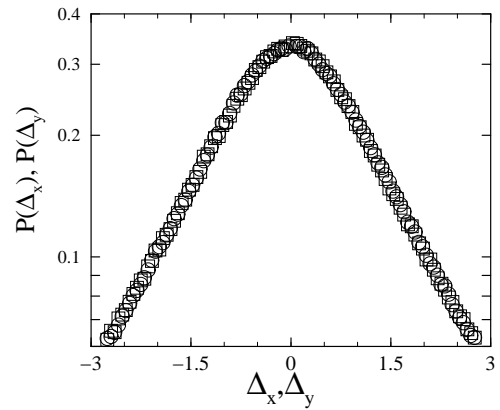}}
\caption{The probability distribution functions of the normalized
  displacements $\Delta_x$ and $\Delta_y$ for all beads of the
  spiral. }
\label{pdf}
\end{figure}

We now focus on the richer fast dynamics. Denoting the displacement in
position between consecutive images for each bead ${\bf \delta} =
(\delta_x,\delta_y)$, we find that they are independent of the bead
position with slight differences for the end beads. We sampled in
excess of $5000$ displacements per bead. The data was obtained in sets
of approximately $540$ frame to frame displacements per experiment
sampled at $250$~Hz. We verified that for each bead the mean value
$\langle {\bf \delta} \rangle$ vanished and that the displacements
were isotropic.  Since the spiral rotates, its fast displacements
sample all directions in the plane so that the orientation of the
coordinate axes is irrelevant. We have plotted the probability
distribution functions $P(\Delta_x)$ and $P(\Delta_y)$ in
Figure~\ref{pdf}. Here, we have defined the normalized displacements
$\Delta_{x,y} = \delta_{x,y}/\sigma_{x,y}$ where $\sigma_{x,y}$ are
the respective standard deviations. Our results show that the
quantities $\Delta_x$ and $\Delta_y$ are identically distributed with
significant incidence for displacements as large as $\pm 3
\sigma_{x,y}$. The probability distribution functions are symmetric
and non-Gaussian. Over populated (with respect to a Gaussian) velocity
distributions \cite{rouyer_00} and, in particular, exponential tails
\cite{olafsen_99,nie_00} are actually observed in excited granular
gases and are understood to be a consequence of the dissipative
bead-bead and bead-plate collisions. Our observations suggest that
velocity statistics of beads in granular chains may be anomalous as
well. Although the exponential tail of the displacement distribution
suggests that the velocity distribution has a stretched exponential
tail, we are unable to discern its precise form due to insufficient
time resolution. 

\vspace{-3mm}

\begin{figure}
\narrowtext
\epsfxsize=6.1cm
\centerline{\epsffile{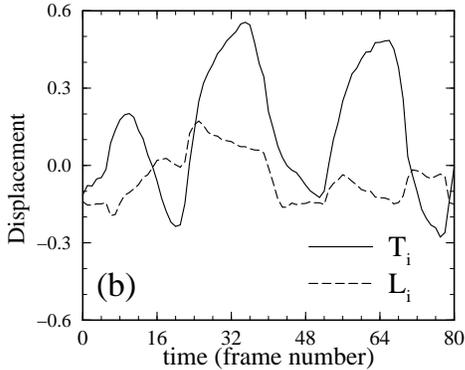}}
\epsfxsize=6.1cm
\centerline{\epsffile{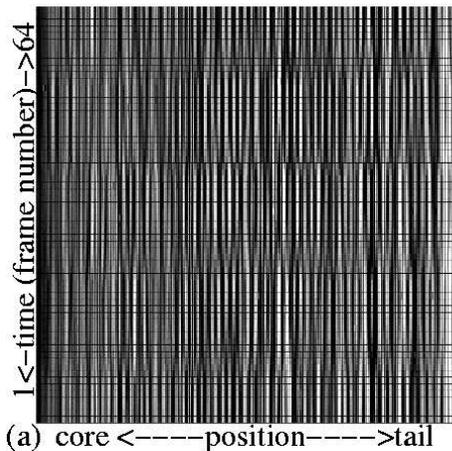}}
\caption{A space-time plot (a) of the transverse displacements of the
  spiral. Position (horizontal axis) along the spiral backbone is
  ordered from core to tail and time (vertical axis) spans $64$
  consecutive images corresponding to 4 drive cycles. Black (white)
  denote a positive (negative) displacement. In (b)
  $T_{i}(t)$ (solid line) and  $L_{i}(t)$ (dashed line) are plotted
  for $i=64$. }
\label{transverse}
\end{figure}

At present we only observe the lateral fast dynamics. The transverse
wave modes illustrated in Figure~\ref{dynamics}b also appear in the
vertical plane. Although the in-plane motion consists primarily of
transverse modes, they must be accompanied by longitudinal modes that
allow for the stretching and contraction of the spiral.  To further
study these modes we consider the locus of the spiral given by the
bead positions ${\bf x}_i = (x_i,y_i)$ and inter-connecting rods ${\bf
r}_i = {\bf x}_{i+1} - {\bf x}_i$.  The mean locus of the spiral
$\langle {\bf r} \rangle_i$ is defined as the time average of ${\bf
r}_i(t)$ over an experimental run. For the frame acquired at time $t$,
we calculate the spiral displacements $\Delta {\bf r}_i (t) = {\bf
r}_i (t) - \langle {\bf r} \rangle_i$. The longitudinal $L_i(t)$ and
transverse $T_i(t)$ displacements from the mean spiral are defined by
$\Delta {\bf r}_i (t) \cdot \langle {\bf r} \rangle_i/|\langle {\bf r}
\rangle_i|$ and $\Delta {\bf r}_i (t) \times \langle {\bf r}
\rangle_i/|\langle {\bf r} \rangle_i|$, respectively.  In
Figure~\ref{transverse}a we have plotted $T_i(t)$ as a function of
position and time. The space-time plot shows that between collisions
with the plate, positive (negative) displacements continue as positive
(negative) {\em i.e.} the beads are roughly ballistic between
collisions. In Figure~\ref{transverse}b we plot representative
displacements for $80$ oscillations of the plate. We see that the
transverse displacements are roughly periodic and are significantly
larger than the longitudinal ones. This behavior is typical for all
inter-connecting rods.

\vspace{-3mm}

\section{Conclusion}
\label{conclusion}

A general expectation from previous experimental work on granular
chains\cite{ebn_01,hastings_02,DRBE_02,prentis_02} is that
configurationally the chain assumes very many different states and in
this manner explores a large region of its phase space. Contrary to
this, we found that a long chain would spontaneously nucleate into a
spiral in a narrow range of weak accelerations.  It is surprising that
a linear chain initiated in a random configuration invariably evolves
to a highly specialized state that occupies a very small volume of the
available phase space. Moreover, it is interesting that minor
differences between the two ends of the chain have such a pronounced
effect.

The origins of spiral formation in vibrated chains is the slight
forward motion of the chain induced by variations at the ends of the
chain. As a result of this asymmetry, the chain crawls forward until
it encounters itself. It then has two choices: to turn inwards and
wrap itself into a spiral or outwards forming locally aligned chain
segments. In the latter case, the end eventually escapes.  With
successive encounters, it becomes likely that the end will be captured
and coil up into a spiral.  The rate of nucleation depends, therefore,
on the forward ratcheting motion coupled with the possibility of
becoming trapped. Such a mechanism would apply to any flexible
forward-translating object confined to two-dimensions and moving
stochastically.

Once formed, the spiral geometry and dynamics is very reproducible.
The spiral becomes looser away from the core and eventually, its tail
is practically linear.  Dynamically, bead-bead and bead-plate
collisions induce transverse and longitudinal modes, the former being
stronger.  On a longer time scale, the spiral rotates in the direction
of coiling and so remains stable.

It will be interesting to further study the spiral formation
mechanism. This may be done for example by manipulating one of the
chain ends using say a relatively larger bead. Observing the rotation
rate as a function of the bead size may elucidate the physical nature
of the ratcheting mechanism. Moreover, the nucleation process is in
itself intriguing. The distribution of waiting times and its extremal
characteristics may alternatively shed light on the same question.

We thank Matthew Hastings, Charles Reichhardt, and James Glazier for
useful discussions. This research is supported by the U.S. DOE
(W-7405-ENG-36).

\end{multicols}
\end{document}